# Feature-based SpMV Performance Analysis on Contemporary Devices

Panagiotis Mpakos*, Dimitrios Galanopoulos*, Petros Anastasiadis*,
Nikela Papadopoulou†, Nectarios Koziris*, Georgios Goumas*

*Computing Systems Laboratory, National Technical University of Athens, Greece
†Chalmers University of Technology, Göteborg, Sweden
*{pmpakos,dgal,panastas,nkoziris,goumas}@cslab.ece.ntua.gr  †nikela@chalmers.se

*Abstract*—The SpMV kernel is characterized by high performance variation per input matrix and computing platform. While GPUs were considered State-of-the-Art for SpMV, with the emergence of advanced multicore CPUs and low-power FPGA accelerators, we need to revisit its performance and energy efficiency. This paper provides a high-level SpMV performance analysis based on structural features of matrices related to common bottlenecks of memory-bandwidth intensity, low ILP, load imbalance and memory latency overheads. Towards this, we create a wide artificial matrix dataset that spans these features and study the performance of different storage formats in nine modern HPC platforms; five CPUs, three GPUs and an FPGA. After validating our proposed methodology using real-world matrices, we analyze our extensive experimental results and draw key insights on the competitiveness of different target architectures for SpMV and the impact of each feature/bottleneck on its performance.

## I. INTRODUCTION

Sparse Matrix-Vector (SpMV) multiplication is a key computational kernel used in a broad range of applications from scientific computing, graph processing, machine learning, and others. SpMV is at the heart of large sparse system solvers, actually dominating their execution time. It has been repeatedly reported to achieve a very small fraction of system peak performance and very high variation across different matrices. A mix of problems like memory bandwidth intensity, reduced ILP, imbalance and memory latency overheads cause this problematic performance behavior [1], [2]. The exact bottleneck that impacts the kernel's performance is influenced by the matrix properties and may differentiate between target architectures.

Traditionally, research effort has focused on the analysis of SpMV performance and the proposal of effective format selection methods [2] that are based on the input matrix [3]–[11], the target platform [12]–[15], or both [2], [16], [17]. In the past few years, GPUs have dominated the field as the state-of-the-art target to accelerate SpMV, due to their high computational throughput and memory bandwidth. Although SpMV creates indirect memory accesses that may lead to divergence in the control flow, GPU-accelerated SpMV has been extensively studied and optimized [18]–[22]. However, with the emergence of new CPUs like AMD EPYC and ARM NEON, with higher core counts and increasing cache sizes, CPUs are becoming compelling again. Furthermore, the requirement for energy-efficient high-performance computing leads to the consideration of energy-related metrics when assessing the effectiveness of an approach. Therefore, emerging architectures such as FPGAs that offer better performance per watt, are now considered viable alternatives for large-scale systems.

The goal of this paper is to study the execution behavior of SpMV in terms of performance and energy efficiency on contemporary devices, considering the four identified potential bottlenecks, i.e., memory bandwidth intensity, reduced ILP, load imbalance and memory latency overheads. To understand the impact of these bottlenecks, we focus on four dominant characteristics of the sparse matrix, which are directly linked to at least one bottleneck (Section III-A). Specifically, we consider the *matrix memory footprint* for memory bandwidth intensity, the *average row length* for low instruction-level parallelism, the *coefficient of skewness* for load imbalance and *irregularity* described by *cross-row similarity* and the *average number of neighbors* between rows for memory latency overheads.

The main contributions of this work are:
- We identify a minimal matrix feature set that adequately captures the dominant performance bottlenecks of SpMV. (Section III-A)
- We implement and make publicly available[1] a sparse matrix generator (Section III-B), that uses the feature set under consideration to generate an artificial matrix dataset, much larger than other commonly used real-world matrix suites and spanning over a much wider space of matrix features.
- We collect experimental results using a multitude of 'state-of-practice' and 'research' storage formats and implementations[2] on nine target devices (Section IV) and validate that despite the minimal feature set used, the artificial matrices exhibit similar performance to real matrices with similar features across all devices (Section V-A).
- Analyzing a wide set of evaluation results, we find that although GPUs maintain their performance supremacy, CPUs are back in the game for medium sized matrices (Section V-B). FPGAs are the most energy-efficient device for SpMV, but with a significantly lower performance ceiling at the moment.
- In our feature-centric analysis (Section V-C), we pinpoint the memory footprint as the most impactful feature, followed by the average number of nonzeros per row. Imbalance is handled by most formats across all devices, while irregularity can imperil GPU performance.
- We analyze performance of 'research' formats (Section V-D) and observe that, while 'state-of-practice' formats have better overall performance, 'research' formats offer a competitive advantage for problematic cases, i.e. at large-size matrices that are unbalanced and irregular.
- We examine the impact of the size of the artificial matrix dataset on the quality of the performance results (Section V-E) and we showcase how artificially-generated datasets can drive data analysis to better understand performance bottlenecks of SpMV (Section V-F).

---

[1]The generator is available at https://github.com/DimitrisGalanopoulos/artificial-matrix-generator
[2]The benchmarks are available at https://github.com/p-anastas/SpMV-Research

## II. BACKGROUND

In this section we discuss common SpMV bottlenecks and storage formats for sparse matrices.

### A. Common SpMV performance bottlenecks

SpMV is renowned for achieving a very low percentage of peak performance, which, even in best cases, may not exceed 20% [13], [21], [23]. Although many factors simultaneously contribute to this low performance, including the matrix sparsity pattern, the storage format, the algorithm, and the hardware characteristics of the platform, prior work [2], [21], [24] converges to the following four identifiable bottlenecks in the performance of SpMV, evident across different architectures:

*1) Memory bandwidth intensity:* SpMV is a kernel with very low operational intensity; its flop-per-byte ratio is lower than 1, due to its streaming nature and the access to the metadata of the sparse storage format. As a result, SpMV is a victim of the processor/memory performance gap. The severity of this bottleneck can vary depending on the size of the input matrix and the available memory bandwidth of the platform, however, it is considered the major bottleneck of the kernel.

*2) Low ILP:* ILP becomes an additional bottleneck when the sparse matrix is populated by rows with a small number of nonzero elements, and loop traversal operations become a significant percentage of the execution time. In these cases, the kernel experiences even lower operational intensity, due to the higher contribution of the vectors and indexing metadata (e.g. row pointers). At the same time, the opportunities for vectorization and row-level parallelism become limited.

*3) Load imbalance:* This bottleneck appears in matrices with high variance in the number of nonzero elements per row or different regions of sparsity patterns. Depending on the work distribution policy, SpMV may suffer from serious load imbalance and resource under-utilization. The severity of this bottleneck depends on the sparsity pattern of the input matrix, the storage format, the algorithm and the work distribution, as well as the available parallelism on the hardware platform.

*4) Memory latency overheads:* This bottleneck plunders SpMV implementations due to the irregular access pattern to the $x$ vector, dictated by the sparsity pattern of the matrix. These accesses can create a significant number of cache misses on CPUs, unaligned memory accesses on GPUs, and significant overheads on FPGAs, if the $x$ vector is not cached on the fastest memory, translating to performance penalties. The impact of these overheads depends on the sparsity pattern of the matrix.

### B. Sparse Matrix Storage formats and Implementations

The matrix in SpMV is stored in a *sparse storage format*, which significantly influences its performance. Various formats have been implemented throughout the years, targeting specific architectural, algorithmic, or sparsity-specific bottlenecks. The plethora of related work on the selection of the optimal storage format [3]–[11] illustrates that no storage format is perfect. In our survey, we consider two format categories; the 'state-of-practice' formats of vendor libraries, and some 'research' formats proposed by academic research. Below we give a basic overview of the strengths and weaknesses of each format used; for more details and pseudocode for the state-of-practice formats we refer the reader to Filippone et al. [21]. We note also that an exhaustive inclusion of all research formats is prohibitive due to time/space limitations and code unavailability. This would also be beyond the scope of this work, which focuses on assessing the performance improvements brought to SpMV by sophisticated storage/optimization techniques and not on an extensive comparison of storage formats.

*1) Coordinate Format (COO):* COO is the straightforward storage format; it stores the sparse matrix in three arrays of length $nnz$ containing each non-zero's row ($RowIdx$), column ($ColIdx$) and value ($Value$). COO performs well in balancing the load among workers, but the redundant metadata information increases the memory bandwidth requirements.

*2) Compressed Sparse Row Format (CSR):* CSR is the most widely adopted format, which compresses $RowIdx$ to a $RowPtr$ an array of size $m+1$, containing the starting element offset for each row. It significantly reduces the matrix's memory footprint and depending on the matrix structure it can achieve some balance between the memory intensity of SpMV and the potential of ILP.

*3) ELLPACK (ELL) and Hybrid (HYB) Formats:* Designed for better utilization of vector units, the ELL format stores the nonzero columns and values in dense $m \times max(nnz\_row)$ arrays $Offset$ and $Evalue$ respectively, adding padding for all smaller rows. This way it improves ILP at the expense of memory bandwidth, achieving high performance for balanced matrices but suffering serious performance degradation for matrices with large variance in non-zeroes per row due to excessive padding. The HYB format aims to avoid this and improve ELL's memory footprint by combining it with COO. HYB defines a threshold $k$, either via heuristics or set to the average number of nonzeroes per row and uses ELL for the first $k$ nonzeroes per row and COO for the remaining elements.

*4) VSL (CSC [21] variant):* VSL [25] is Xilinx's specialized format for the Alveo-U280 FPGA. It splits the matrix in 2D

partitions which in turn are divided in 16 parts and fed to 16 execution units by equal HBM channels, using zero-padding in order to accommodate for the double-precision accumulation latency. This design fails when excessive padding is applied and the storage requirements of the matrix exceed the maximum capacity of the HBM channels utilized to store the matrix.

*5) Research formats and implementations:* CSR5 [20] extends CSR to tackle load imbalance, by splitting the rows to create similar-sized subproblems. It is designed to fit GPUs, where CSR fails to balance work between warps/threads. The drawback of CSR5 is its requirement for additional metadata for row splitting, which slightly increases memory footprint, however it can result to slowdowns in cases where its sophisticated splitting of the input matrix is fruitless. Merge [26] is another lightweight extension of CSR, with no preprocessing cost. It overcomes load imbalance, by assigning equally-sized chunks of work to each processing element. SELL-C-$\sigma$ [27] is a variant of ELL, where the tunable parameters *C* (chunk size) and $\sigma$ (sorting scope) are selected to match the underlying hardware capabilities without increasing memory latency overheads. SparseX [28] is a library that automatically detects dense, horizontal, vertical, diagonal or block substructures in a sparse matrix. SparseX directly targets memory bandwidth intensity and encodes each substructure with a minimal memory footprint, resulting in a highly compressed representation of the matrix, something that can be beneficial especially for large matrices.

## III. METHODOLOGY

The goal of this study is to gain insights on the performance behavior of SpMV on contemporary compute devices. In order to achieve this, we need to collect evaluation results from a large number of matrices, making sure that the aforementioned potential bottlenecks do appear in our experiments. For this, we choose not to rely on real matrices that already exist in suites [29], [30], since a) they are not sufficient in number, b) many of them are correlated (e.g. different versions of the same matrix) and c) they are not always able to capture performance behaviors, even if such behaviors occur in rare matrices, or even in matrices that do not (currently) exist.

In order to control the experimental process, we resort to the generation of artificial matrices. A key step in our methodology is to decide upon the matrix creation process. Our approach relies on the definition of one (or a very limited number of) matrix feature(s) that is highly correlated with at least one of the performance bottlenecks discussed in Section II-A. In the next section we discuss the five features selected. With the matrix features at hand, we are able to define a feature space bounded by the limits of real-life matrices, and then create a generator that traverses the search space and creates a large number of matrices under our control.

### A. Matrix features

Prior work has identified the correlation of matrix features with the performance of SpMV [2], [6]–[9]. A rather high number of features have been used to train proper predictors for SpMV performance. We draw inspiration from prior work and focus on matrix-related features that influence performance, but we deliberately restrict the number of features, and in particular use only one per performance bottleneck. In this way, we trade accuracy for simplicity, better understanding of the results and more controllable process in matrix creation.

TABLE I: Features used for artificial matrix generation

| label | feature | description | matrix space |
|---|---|---|---|
| f1 | mem_footprint | matrix (CSR) size (MB) | [4-32], [32-512], [512-2048] |
| f2 | avg_nz_row | avg nonzeros per row | 5, 10, 20, 50, 100, 500 |
| f3 | skew_coeff | skewness coefficient | 0, 100, 1000, 10000 |
| f4.a | cross_row_sim | row similarity | 0.05, 0.5, 0.95 |
| f4.b | avg_num_neigh | neighbors number | 0.05, 0.5, 0.95, 1.4, 1.9 |

*1) Matrix memory footprint as an indicator for memory bandwidth intensity:* The memory footprint (in MBytes) of the sparse matrix data in the CSR format has a big impact on performance. Small sizes fit in the lower levels of the memory hierarchy (e.g. cache in a CPU, shared memory in a GPU etc), but also limit the parallel slack. Large sizes offset the overheads of the more complicated formats and can utilize more workers, but also demand more memory bandwidth.

*2) Average nonzeros per row as an indicator of low ILP:* The number of nonzero elements per row clearly influences ILP, as, for small rows, a large percentage of CPU cycles is spent for loop control instructions. However, this metric has also additional severe implications on performance as it reduces the operational intensity, can indicate the density (regularity) of the matrix, the potential for vectorization, and its balance (very large average nonzeros per row cannot easily lead to imbalanced matrices).

*3) Skew coefficient as an indicator of imbalance:* This metric measures how unbalanced the rows of the matrix are, with respect to the number of nonzeros, and is defined as:

$$skew = \frac{<max\ nnz\ per\ row> - <average\ nnz\ per\ row>}{<average\ nnz\ per\ row>}$$

A skew of 1 means that the longest row is twice as big as the average number of nonzeros per row. Typically, matrices with skews around ten or lower are considered balanced; unbalanced matrices have skews ranging in the hundreds or even thousands.

*4) Average number of neighbors and cross-row similarity as an indicator of irregularity and memory latency overheads:* We define as "neighbors" of a nonzero element all the other *same-row* elements residing in a predetermined maximum column distance, left or right of the element. This metric computes the average number of neighbors with a column distance of one (therefore a number between 0 and 2), and it is a measure of nonzero clustering, and spatial locality on vector $x$.

*Cross-row similarity* expresses a measure of similarity between adjacent rows. We define the "cross-row neighbors" of a nonzero element, which are all the elements in the *next* row, that reside at a predetermined maximum column distance. We used a column distance of one, which means either the same column, or the adjacent left and right ones. We then compute the cross row similarity metric as the fraction of a row's elements that have at least one cross row neighbor (a number between 0 and 1), averaged across all rows. Cross-row similarity targets to capture temporal locality on vector $x$.

TABLE II: Testbed Characteristics and the storage formats/libraries used per testbed.

| Testbed | AMD-EPYC 7402 (AMD-EPYC-24) | AMD-EPYC 7742 (AMD-EPYC-64) | ARM-NEON Ampere-Altra Q80-33 | Intel Xeon Gold 5120 |
|---|---|---|---|---|
| Cores | 24 cores | 64 cores | 80 cores | 14 cores |
| Memory | 128 MB LLC (L3)<br>256 GB DDR4 | 256 MB LLC (L3)<br>256 GB DDR4 | 80 MB LLC (L2)<br>512 GB DDR4 | 19.25 MB LLC (L3)<br>256 GB DDR4 |
| Measured BW | DDR4 50 GB/s, LLC 700 GB/s | DDR4 105 GB/s, LLC 878 GB/s | DDR4 102 GB/s, LLC 650 GB/s | DDR4 55 GB/s, LLC 300 GB/s |
| Compiler | gcc 9.2.0 | gcc 9.2.0 | armclang 22.0.1 | gcc 7.5.0 |
| Compiler Flags | -O3 -mavx2 -fopenmp | -O3 -mavx2 -fopenmp | -O3 -fopenmp -armpl=lp64,parallel | -O3 -fopenmp |
| Formats | MKL-IE 2018.1, AOCL-Sparse 3.2,<br>Naive-CSR, Vectorized-CSR, CSR5 [20],<br>MergeCSR [26], SparseX [28], SELL-C-$\sigma$ [27] | MKL-IE 2019.1, Naive-CSR, CSR5 [20] | Arm Perf. Library 21.1.0, Naive-CSR,<br>Naive-CSR, Merge-CSR [26],<br>SparseX [28], SELL-C-$\sigma$ [27] | MKL-IE 2018.1, Naive-CSR, CSR5 [20],<br>Vectorized-CSR, MergeCSR [26],<br>SparseX [28], SELL-C-$\sigma$ [27] |

| Testbed | IBM POWER9 AC922 | NVIDIA Tesla-P100 (PCIe) | NVIDIA Tesla-V100 (PCIe) | NVIDIA Tesla-A100 (PCIe) | Alveo-U280 |
|---|---|---|---|---|---|
| Cores | 16 cores (2 threads/core) | 3584 CUDA cores @ 1480 MHz | 5120 CUDA cores @ 1455 MHz | 6912 CUDA cores @ 1412 MHz | 1.3M LUTs |
| Memory | 80 MB LLC (L3)<br>319 GB DDR4 | 12 GB CoWoS HBM2<br>BW 549 GB/s | 32 GB CoWoS HBM2<br>BW 900 GB/s | 40 GB CoWoS HBM2<br>BW 1555 GB/s | 8 GB HBM2 - BW 460GB/s<br>32 GB DDR4 - BW 38GB/s |
| Measured BW | DDR4 109 GB/s, LLC 612 GB/s | HBM2 BW 464 GB/s | HBM2 BW 760 GB/s | HBM2 BW 1350 GB/s | HBM2 BW 287.5 GB/s |
| Compiler | gcc 8.3.1, IBM XL 16.1.1 | gcc 9.2.0, cuda-9.2 & cuda-11 | gcc 9.2.0, cuda-9.2 & cuda-11 | gcc 9.2.0, cuda-11 | gcc 7.5.0, Vitis 2020.2 |
| Compiler Flags | -O3 -fopenmp | -O3 -arch=sm_60 | -O3 -arch=sm_70 | -O3 -arch=sm_80 | -O3 |
| Formats | Naive-CSR, Balanced-CSR,<br>MergeCSR [26], SparseX [28] | cuSPARSE-11 COO & CSR,<br>cuSPARSE-9.2 HYB, CSR5 [20] | cuSPARSE-11 COO & CSR,<br>cuSPARSE-9.2 HYB, CSR5 [20] | cuSPARSE-11 COO & CSR,<br>MergeCSR [26] | Vitis Sparse library 2021.1 |

Clearly, both features of irregularity are quite simple and brute force, e.g. they do not investigate relations beyond consecutive rows. A more sophisticated scheme (e.g. stack distance profiles) could better capture memory access effects but would also incur unnecessary complexity for the insights that we aim to gain. Our initial attempts to define a single feature for irregularity focused on row bandwidth (or proper splits of it), but would not capture an adequately accurate behavior. We thus resorted to the two metrics for spatial and temporal locality. We still use row bandwidth internally in our generator, as mentioned in Section III-B.

```
csr_matrix * artificial_matrix_generation(
    nr_rows, nr_cols, avg_nz_row, std_nz_row,
    distribution, skew_coeff, bw_scaled,
    cross_row_sim, avg_num_neigh)
```

Listing 1: Artificial matrix generation function

### B. Matrix Generator

After investigating real-life matrices found in literature and matrix suites, we defined the feature space shown in Table I, so that it represents as uniformly as possible the majority of the real matrix types. We then generated artificial matrices covering this feature space, via the matrix generator described below. The generator function is similar to the one in Listing 1 and is quite flexible in defining the matrix features. The function returns the artificial matrix data in the CSR storage format, which we then convert to whichever format is being tested.

The number of rows and the average nonzeros per row implicitly define the size of the matrix. The generator uses a random distribution (given by the `distribution` argument) to determine the number of nonzeros for each row. In this study we utilized the normal distribution: $\mathcal{N}(\text{avg\_nz\_row} ; \text{std\_nz\_row})$.

To achieve the skew between the number of nonzeros of the rows we utilized an exponentially decreasing function, starting with the maximum value needed, i.e. $MAX \cdot e^{\frac{-C \cdot \text{row\_idx}}{\text{nr\_rows}}}$, where $C$ is a calculated constant controlling the shape of the function (used for the standard deviation). The average of the previous distribution function is then recalculated, so that the combined average of nonzeros per row is equal to the desired one.

After selecting the number of nonzeros, we define their position through the remaining arguments. The algorithm works row-wise. As a first step, we duplicate column positions from the previous row, with a probability of `cross_row_sim`, to achieve similarity between adjacent rows. In the last step we use a randomized positioning scheme. The `bw_scaled` argument is the matrix bandwidth, scaled as a fraction of the number of columns ([0-1]). We use $0.05, 0.3$ and $0.6$ for `bw_scaled`. The nonzeros of the row are randomly placed in a confined space in each row, so that the average bandwidth matches the one given. After each random placement we also repeatedly place adjacent neighbors, with a probability derived from `avg_num_neigh`, until we fail the dice roll. This way we can simulate the same-row nonzero clustering.

## IV. EXPERIMENTAL SETUP

In our experiments we evaluate SpMV on three different architectures (a total of nine different machines): three NVIDIA GPUs (Tesla-V100, Tesla-P100, Tesla-A100), five single-socket CPU systems (AMD-EPYC 7402, AMD-EPYC 7742, ARM-NEON Ampere Altra Q80-33, Intel Xeon Gold 5120, IBM Power9 AC922) and a Xilinx Alveo-U280 FPGA. Table II lists in more detail the testbed hardware and formats/libraries utilized. The AMD-EPYC-24 and AMD-EPYC-64 were configured as NPS1 and NPS4 (NUMA nodes per socket) respectively. The IBM CPU supports up to 4 hardware threads per core. After testing, the best overall configuration was identified as 2 threads per core. No special configuration was applied for the other CPUs. The threads were always pinned to the cores using the OpenMP environment variables, and the matrices were initialized in a parallel fashion, utilizing the Linux first-touch policy, in order to ensure data locality. We calculate the memory bandwidth using the STREAM benchmark for all testbeds except the FPGA, where we calculate bandwidth based on the number of utilized HBM2 channels (20 out of 32) Shedding more light to multiple device execution behavior (e.g. dual CPU/socket) is left for future work.

As mentioned in Section II-B, we use publicly available, 'state-of-practice' formats for each device, as well as 'research' formats and implementations, avoiding those that are specialized for a specific matrix family, in order to provide a general analysis. In some cases research formats may also involve pre-processing overheads or tuning complexity. This is left out of our analysis as we focus on pure SpMV performance. We note that, due to access limitations to the AMD-EPYC-64 system, we were not able to run experiments on all formats tested on other CPU systems. However, the general architectural trends were captured adequately with the available formats and they were inline with the smaller AMD-EPYC-24 behavior. Additionally, the range of research formats tested in the Tesla-A100 was limited by the lower availability of CUDA-SDK version 11 updated formats, the minimum requirement for A100's compute capability of 8.0.

Through the matrix features described in Section III-A we define a search space (Table I), and we generate 16200 artificial matrices within it. In each configuration (testbed/matrix/format) we ran 128 iterations of double precision SpMV, and record the average performance (GFLOPs) and power consumption (Watt), taking the arithmetic mean of 5 experiments. We use double precision as the standard data type for scientific computations and leave the study of other precision levels for future work. For AMD and Intel CPUs, we compute the average power consumption by collecting energy readings for the CPU package (core and uncore components) through `RAPL` (Running Average Power Limit), for the duration of the execution. Similarly, we collect energy readings from the `Altra-HWMON` (Hardware Monitor) driver for ARM-NEON. We were not able to collect accurate power consumption measurements on Power9 system. We therefore use a pessimistic estimation of a constant, 200W TDP for the CPU socket. We include these results for the sake of completeness. For the GPUs and FPGA, we estimate the average power by collecting power readings throughout the execution, using the `nvml-driver` interface of the `nvidia-smi` and the `xbutil` tool, respectively.

## V. EXPERIMENTAL EVALUATION

### A. Validation of artificial matrices

Before analyzing SpMV performance based on our artificial matrices, we need to validate that the matrix generation approach is able to capture the behavior of real matrices with a reasonable accuracy. Our basic assumption for validation is that if our approach is effective, then artificial and real matrices with similar characteristics should have similar performance. We form a validation suite composed of the 45 most widely used matrices in prior literature that cover a wide spectrum of the examined features, listed in Table III. For each real matrix, we generate additional artificial matrices (separate from the artificial dataset of Table I), using a wide $[-30\%, +30\%]$ value range for each feature, and we will be referring to these as its artificial '*friends*' (roughly 70 per matrix). We report the best performance achieved among tested formats on each platform per matrix.

Fig. 1 shows the performance of the validation matrices (black dots) and the performance range of their corresponding 'friend' artificial matrices (boxplots) for each testbed. For a better understanding of performance, we also provide an estimation of the Roofline bound [31] using the CSR memory footprint of each validation matrix and the measured bandwidth for each testbed displayed in Table II. We note that 10 matrices fail to execute on the FPGA due to HBM capacity limitations. We also note that we were not able to collect measurements for certain matrices on 3 devices (Tesla-P100, AMD-EPYC-64, ARM-NEON), due to resource unavailability at the time of experimentation. We make the following observations: First, many validation and friend matrices are close to their corresponding roofline bound, providing an evidence that the kernel implementations used are well optimized. The cases that show the highest divergence from the roofline bound are a) very small matrices in GPUs which cannot fully utilize their resources, b) the FPGA testbed in general, where excessive padding complicates roofline performance characterization and c) matrices with problematic features (excluding memory footprint). We refer to a matrix as 'problematic' when its bad performance cannot be explained by its memory footprint alone, i.e., it is not memory-bandwidth bound, but from a combination of other bottlenecks (low ILP, imbalance, irregularity) that cannot be interpreted by the memory roofline model.

TABLE III: Matrix suite used for validation. The range of each regularity (f4) subfeature is split in 3 equal subranges, with labels S(mall), M(edium), L(arge) ("Small" implies an irregular matrix).

| id | Matrix | f1 | f2 | f3 | f4 |
|---|---|---|---|---|---|
| 1 | scircuit | 11.63 | 5.61 | 61.95 | MM |
| 2 | mac_econ_fwd500 | 15.36 | 6.17 | 6.14 | MS |
| 3 | raefsky3 | 17.12 | 70.22 | 0.14 | LL |
| 4 | bbmat | 20.42 | 45.73 | 1.76 | LM |
| 5 | conf5_4-8x8-15 | 22.13 | 39 | 0 | LL |
| 6 | mc2depi | 26.04 | 3.99 | 0 | LS |
| 7 | rma10 | 27.35 | 50.69 | 1.86 | LL |
| 8 | cop20k_A | 30.5 | 21.65 | 2.74 | MM |
| 9 | thermomech_dK | 33.35 | 13.93 | 0.44 | MM |
| 10 | webbase-1M | 39.35 | 3.11 | 1512.43 | LS |
| 11 | cant | 46.1 | 64.17 | 0.22 | LL |
| 12 | ASIC_680k | 46.91 | 5.67 | 69710.56 | LM |
| 13 | pdb1HYS | 49.86 | 119.31 | 0.71 | LL |
| 14 | TSOPF_RS_b300_c3 | 50.67 | 104.74 | 1 | LL |
| 15 | Chebyshev4 | 61.8 | 78.94 | 861.9 | LL |
| 16 | consph | 69.1 | 72.13 | 0.12 | LL |
| 17 | com-Youtube | 72.71 | 5.27 | 5460.3 | MS |
| 18 | rajat30 | 73.13 | 9.59 | 47421.8 | MM |
| 19 | radiation | 88.26 | 34.23 | 101.18 | SS |
| 20 | Stanford_Berkeley | 89.39 | 11.1 | 7519.69 | MM |
| 21 | shipsec1 | 89.95 | 55.46 | 0.84 | LL |
| 22 | PR02R | 94.29 | 50.82 | 0.81 | LM |
| 23 | gupta3 | 106.76 | 555.53 | 25.41 | LL |
| 24 | mip1 | 118.73 | 155.77 | 425.24 | LL |
| 25 | rail4284 | 129.15 | 2633.99 | 20.33 | SL |
| 26 | pwtk | 133.98 | 53.39 | 2.37 | LL |
| 27 | crankseg_2 | 162.16 | 221.64 | 14.44 | LL |
| 28 | Si41Ge41H72 | 172.5 | 80.86 | 7.19 | LM |
| 29 | TSOPF_RS_b2383 | 185.21 | 424.22 | 1.32 | LL |
| 30 | in-2004 | 198.88 | 12.23 | 632.78 | LL |
| 31 | Ga41As41H72 | 212.61 | 68.96 | 9.18 | LM |
| 32 | eu-2005 | 223.42 | 22.3 | 312.27 | LM |
| 33 | wikipedia-20051105 | 232.29 | 12.08 | 410.37 | SS |
| 34 | human_gene1 | 282.41 | 1107.11 | 6.17 | SS |
| 35 | delaunay_n22 | 304 | 6 | 2.83 | MS |
| 36 | sx-stackoverflow | 424.58 | 13.93 | 2738.46 | SS |
| 37 | dgreen | 442.43 | 31.87 | 4.87 | SS |
| 38 | mawi_201512012345 | 506.18 | 2.05 | 8006372.09 | LM |
| 39 | ldoor | 536.04 | 48.86 | 0.58 | LL |
| 40 | dielFilterV2real | 559.9 | 41.94 | 1.62 | MM |
| 41 | circuit5M | 702.4 | 10.71 | 120504.05 | LM |
| 42 | soc-LiveJournal1 | 808.06 | 14.23 | 1424.81 | SS |
| 43 | bone010 | 823.92 | 72.63 | 0.12 | LL |
| 44 | audikw_1 | 892.25 | 82.28 | 3.19 | LL |
| 45 | cage15 | 1154.91 | 19.24 | 1.44 | LS |

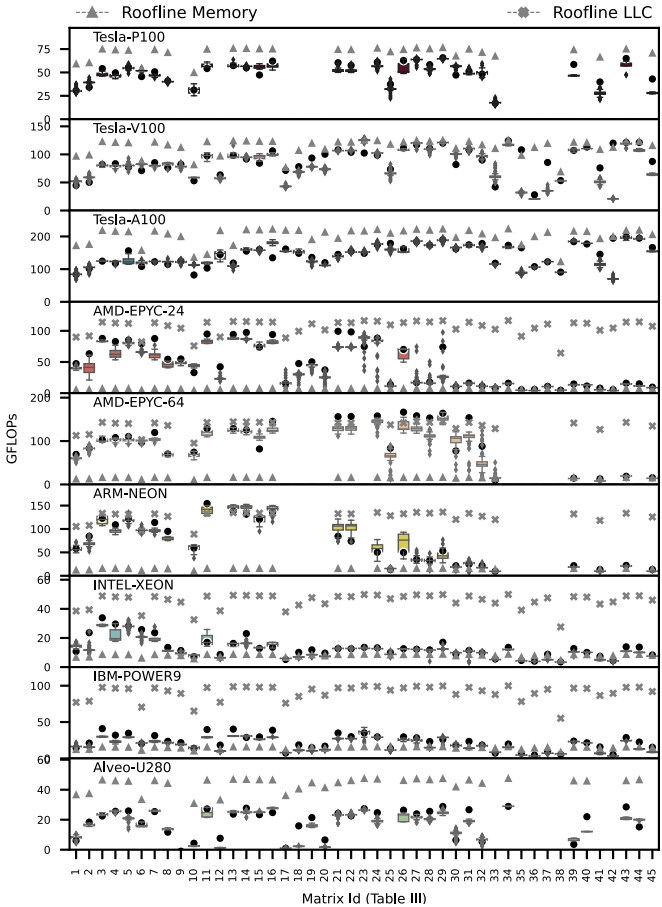

Fig. 1: Performance Comparison between the selected Validation matrices of Table III and their 'friends' in our testbeds. The Roofline Memory points (--△--) denote the memory bandwidth performance roof for each validation matrix, and the Roofline LLC points (--X--) the respective last-level cache roof.

TABLE IV: Comparison between the performance of each validation matrix and 1) the performance of its synthetic 'friends' (Mean Absolute Percentage Error - MAPE) or 2) the performance of its closest 'friend' (APE-best)

| Device | MAPE | APE-best |
|---|---|---|
| Tesla-P100 | 10.01 % | 4.57 % |
| Tesla-V100 | 18.42 % | 10.15 % |
| Tesla-A100 | 9.94 % | 5.19 % |
| AMD-EPYC-24 | 20.04 % | 8.42 % |
| AMD-EPYC-64 | 21.81 % | 6.39 % |
| ARM-NEON | 15.65 % | 4.41 % |
| INTEL-XEON | 16.49 % | 7.36 % |
| IBM-POWER9 | 21.77 % | 14.11 % |
| Alveo-U280 | 23.49 % | 16.63 % |
| **Average** | **17.51 %** | **8.58 %** |

Second, the artificial matrices in the vast majority of the cases are able to follow the performance trend of real matrices in all devices. Specifically, the mean absolute percentage error

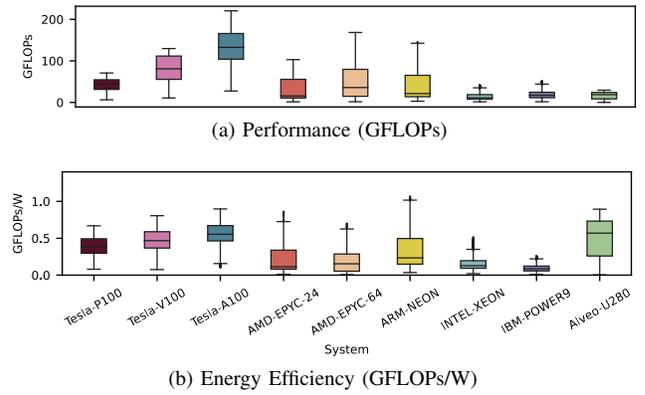

(a) Performance (GFLOPs)

(b) Energy Efficiency (GFLOPs/W)

Fig. 2: Performance and Energy Efficiency of SpMV on different platforms.

(MAPE) between the performance of each validation matrix and the median performance of its friends across all matrices and testbeds is $17.51\%$. Third, there are some problematic validation matrices areas that contribute most to the MAPE, especially for the CPU systems, e.g. matrices 30 to 33 for AMD-EPYC-64. This corresponding area for each testbed starts when the validation matrix memory footprint exceeds the cache size leading in performance degradation ('move' from LLC to memory roofline bound), where the $[-30\%, +30\%]$ feature search range can lead to friends fitting in cache while their validation base does not, and vice versa, increasing MAPE. Finally, even validation matrices that differ considerably from their friends' median tend to have very similar performance with at least some of their artificial friends. Specifically, defining as "best friend" for each validation matrix its closest-performing friend, the MAPE between the performance of each validation matrix and their best friend is only $8.58\%$.

Takeaway 1: *The small feature set including the memory footprint, average row size, skew coefficient, average number of neighbors and cross-row similarity is able to capture the performance behavior of sparse matrices across all devices with remarkable accuracy.*

### B. Cross-device SpMV behavior

In this section, we evaluate the performance and energy efficiency of SpMV in all platforms using our artificial matrix dataset. Again, we present the best result achieved among tested formats for each matrix.

*1) Performance:* Fig. 2a summarizes the performance (throughput) for each platform. Our main observations are as follows: a) SpMV performance varies dramatically in our wide dataset of artificial matrices in all devices, b) Although A100 exhibits better performance on average, the latest generation CPUs (AMD-EPYC-64, ARM-NEON), with increased core count and larger caches, emerge as a solid alternative for a number of cases, where high performance is achieved, c) Alveo U280 is not able to compete the throughput of other devices.

Takeaway 2: *CPUs cannot dethrone GPUs in terms of performance, but emerge as a solid alternative.*

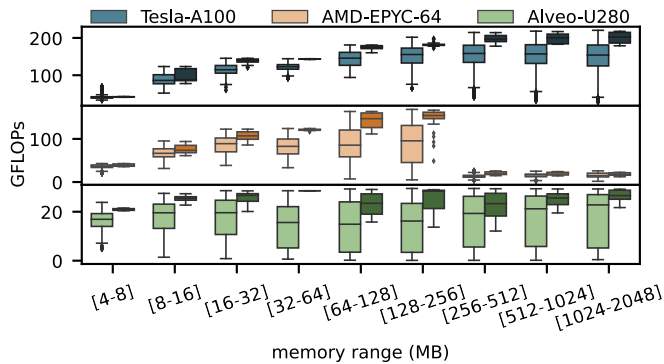
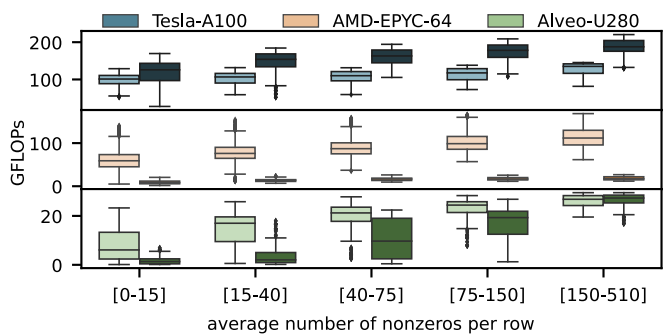

Fig. 3: Impact of memory footprint on SpMV performance. Light-colored boxplots refer to the complete matrix dataset, while dark refer to per-device favorable-featured matrices.

Fig. 4: Impact of row size on SpMV performance. Light colored boxplots refer to small matrices, while dark refer to large matrices. The split threshold is set at 256MB for all devices.

*2) Energy Efficiency:* Apart from pure performance, we also focus on the energy efficiency of each platform tested. Energy-efficiency can be defined as the ratio of useful work performed by a system over a period of time and the energy consumed by the system during this period [32]. In our analysis, we use the GFLOPs/W metric and present results in Fig. 2b. We observe that FPGAs, despite being weaker in terms of raw performance, prevail as the most energy efficient of the tested platforms, operating at a lower power level compared to other devices. GPUs keep a competitive position in energy efficiency, something that can be accredited to the high performance levels that they reach. The CPUs seem to fall significantly behind in this metric, with the exception of ARM-NEON, which was the only CPU to stand out in terms of power consumption.

Takeaway 3: *The three more energy-efficient devices follow three different paths towards energy efficiency: the 'low-power' (Alveo-U280) path, the 'high-performance' (Tesla-A100) path and the 'balanced architecture' path (ARM-NEON).*

### C. Feature Analysis

In this subsection, for simplicity and due to space limitation, we analyze results for one GPU device (Tesla-A100), one CPU device (AMD-EPYC-64) and the only available FPGA. We visualize our experimental results feature-wise, with a goal to gain better understanding on the impact of each on the performance of SpMV per device. Therefore, we choose to show the most powerful GPU and CPU, despite the relatively smaller set of storage formats available on these devices.

*1) The impact of memory footprint (matrix size):* The impact of memory footprint (in MB) on performance is presented in Fig. 3. We provide two boxplots per device, one including all measurements (light-colored boxplot) and one including only the matrices in which the other three features are selected to be intuitively favorable (i.e., regular, balanced matrices with long rows). Clearly, the memory footprint (matrix size) is an extremely impactful feature on GPUs and CPUs, but shows a rather random behavior on FPGAs.

In the case of A100, the device is gradually taking advantage of the additional parallelism existing in larger matrices, and thus favors large over small matrices. We notice that the performance gap between small and large matrices can exceed $2\times$. This behavior is attributed to the extremely high parallelism available in modern GPUs which requires sufficient work to keep thousands of execution units busy. We note also that in a few cases of the large matrices, the performance can collapse to a level much lower than small matrices, revealing an infrequent but major impact of the other three features. Notably, in the case of GPUs, the matrix size does not affect memory bandwidth intensity, it rather affects the levels of available parallelism.

The general performance trend of CPUs is rather expected. Matrices that fit the LLC excel in performance. On the contrary, due to the extreme memory bandwidth intensity of the kernel, the performance for larger matrices collapses, exhibiting a performance gap between small and large matrices that exceeds $7\times$ for the AMD-EPYC-64 CPU. We need to note though that since the LLC size of CPUs is quite large, contemporary CPUs seem to be able to reach high performance for moderately sized matrices, i.e., with 10-20M nonzero elements and up to 1M rows. The other CPUs that were tested show similar behavior, as the memory footprint of the matrix grows, but the different LLC size determines where the performance 'cutoff' point (above 256 MB for AMD-EPYC-64) for each device is.

As mentioned previously, every examined CPU cannot reach the levels of performance of the Tesla-A100 GPU. However, if we focus on matrices whose memory footprint lie in the range 64-256MB, we observe that the AMD-EPYC-64 CPU is able to reach 60% of the Tesla-A100 performance. Similarly for the ARM-NEON, focusing on the matrices that lie in the 32-128MB range, 72% of the A100 performance can be achieved.

Takeaway 4 : (follow-up on Takeaway 2) *CPUs in their favorable size of operation verge on the performance levels of a high-end GPU.*

The performance analysis for FPGAs is more complex. We note that, due to limitations of the Vitis Sparse Library and the underlying hardware, not all matrices in our dataset could run on the Alveo FPGA. Specifically, matrices with many rows, that are highly sparse (few nonzeros per row) and irregular resulted in extensive zero-padding and in memory footprint greater than the capacity of the HBM channels utilized. Therefore, the majority of the large size (greater than 256 MB) matrices that are shown in Fig. 3 exhibit better and more stable performance (narrower boxplots) on the FPGA, being intuitively favorable matrices, compared to small ones. Overall, we observe that the memory footprint causes a rather random behavior, and we need to investigate the other features to get a better insight for SpMV performance on the FPGA.

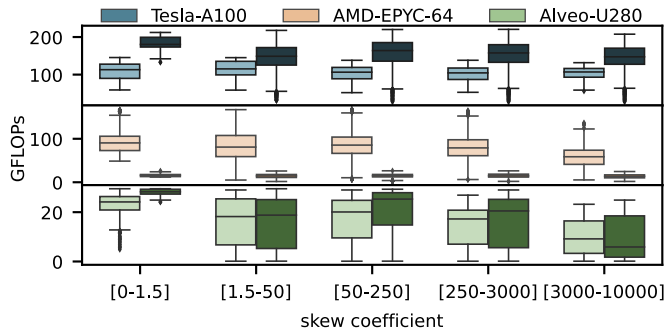
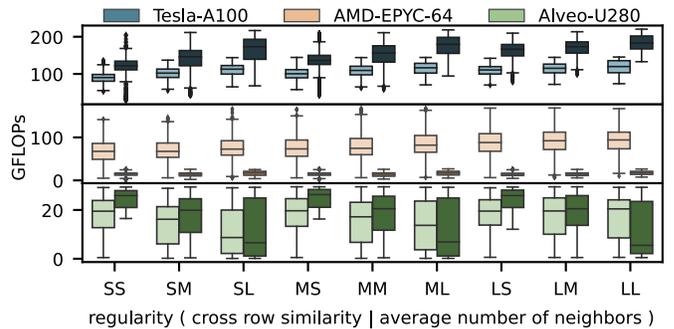

Fig. 5: Impact of imbalance on SpMV performance. Higher skewness means more unbalanced matrix. Light colored boxplots refer to small matrices, while dark refer to large matrices. The split threshold is set at 256MB for all devices.

Fig. 6: Impact of regularity on SpMV performance. Values for subfeatures are set at S(mall), M(edium), L(arge). Higher values refer to more regular matrices. Light-colored boxplots refer to small matrices, while dark refer to large matrices. The split threshold is set at 256MB for all devices.

*2) Row size:* Fig. 4 shows the impact of row size on the performance of SpMV. Based on the discussion in previous paragraph, since the behavior of CPUs and GPUs largely varies between matrix size areas, we split our results for large and small matrices (different split per device). We follow the same approach for FPGAs as well to investigate any interesting patterns that may arise in this case by the combination of matrix and row sizes. As an initial observation, it is clear that this feature is also impactful to performance, however its impact on the GPU and CPU is higher in the favorable matrix sizes for each device (small for the CPU and large for the GPU). Indeed, looking at the medians of the large and small matrix sizes for the GPU the CPU respectively, we observe that the gap between small and large rows is around $2\times$.

On the FPGA, for large-size matrices, the performance degradation due to small row lengths is $\sim 20\times$, verifying our previous statement that this library cannot handle highly-sparse large size matrices well, due to zero-padding. For small-size matrices, the gap between small and large rows is $\sim 7.5\times$. In both cases, matrices with large rows reach the peak performance than can be achieved with this format on this platform.

*3) Imbalance:* The impact of imbalance on SpMV performance is shown in Fig. 5. As in *row size*, we split the results for large and small matrices, with a different split threshold per device. Increasing the coefficient of skewness leads to a more unbalanced matrix, with a small subset of the rows being relatively longer compared to the rest. We visualize the results separately for small and large matrices.

No significant difference is observed between balanced and unbalanced matrices, for the GPU and the CPU. The CPU is less prone to this feature, mainly due to its lower parallelism, as the matrix is distributed to the threads in larger chunks that can more easily hide the imbalance. Most GPU formats on the other hand are designed with work sharing and imbalance in mind, due to the massive parallelism (e.g. zero padding in ELL, splitting rows in CSR5, COO in HYB). For the GPU, highly balanced matrices lead to only $1.2\times$ higher performance. FPGA performance, on the contrary, drops by $\sim 4\times$, and large size matrices suffer more from this performance degradation.

*4) Irregularity:* The last feature that we examine is the combination of intra-row neighbors (`avg_num_neighbors`) and similarity between adjacent rows (`cross_row_similarity`), as indicators of the matrix irregularity. Same as before, we split results for small and large matrices. In Fig.6, we split the range of each subfeature into three equal subranges and label each subfeature as *S(mall)*, *M(edium)*, *L(arge)*, with *"Small"* implying an irregular matrix.

Focusing on the favorable matrix size per platform, we first observe that the GPU is prone to irregularity when running large size matrices. The more regular the matrix, the more robust the performance (boxplot shrinks upwards). In the CPU, the impact of irregularity is lower, and the performance improves by $\sim 1.3\times$ when a matrix becomes regular. We attribute this to the fact that for small/medium sized matrices that fit in the cache, the $x$ vector is also expected to fit in the cache, thus not suffering from cache misses. For larger matrices, the additional penalty paid to cache misses on $x$ seems minor compared to memory bandwidth constraints. Finally for the FPGA, we observe some unexpected behavior. When `avg_num_neighbors` is set to *"Large"*, performance drops, regardless of the value of `cross_row_similarity`. A possible explanation for this is that when `avg_num_neighbors` is set to *"Large"*, more matrices could be generated, due to less zero padding applied. Therefore, the boxplots of *"Large"* `avg_num_neighbors` contain many large size matrices, that do not perform well on the FPGA.

Takeaway 5: *Our observations can be summarized as follows:*

- *For the CPUs, the most impactful feature is the memory footprint of the matrix. When exceeding its cache size, performance can drop by $7\times$. The next most important feature is the row length, with small row lengths decreasing performance by $\sim 2\times$. Imbalance is handled sufficiently, since the level of parallelism is not that high in the CPU,*

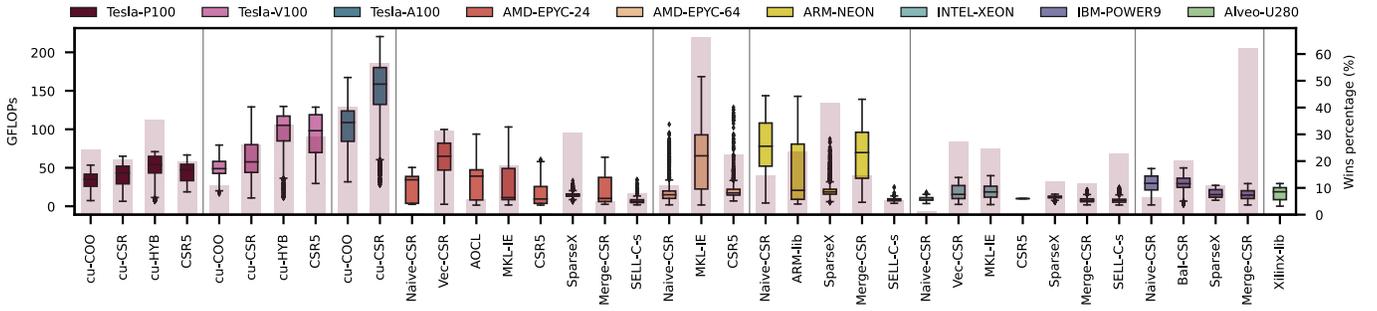

Fig. 7: Performance comparison of SpMV formats/libraries on different devices. The bar behind each format represents the percentage of the artificial matrix dataset that runs optimally with that format for each device separately. 'Naive-CSR' is the standard CSR, 'Bal-CSR' adds nonzero balancing (row resolution) and 'Vec-CSR' also adds vectorization of operations.

*and irregularity does not pose a major problem either.*

- The GPUs have no problem with large matrices and performance is up to 2×, compared to small size ones. Similar to the CPUs, the row length drops performance by 2x when small. Imbalance does not affect performance, with specialized work sharing of best-performing formats. However, irregularity is a problem for GPUs, since performance of large matrices (the favorable size for GPUs) drops considerably, up to 2×.
- For the FPGA, the matrix size does not affect performance. On the other hand, the row length is crucial, since the library cannot handle highly-sparse large size matrices well, due to extensive zero-padding. Imbalance drops performance by up to ∼4×, especially for large size matrices. Regarding irregularity, an unusual behavior is observed, due to extensive padding and limited memory capacity.

### D. Impact of matrix formats

In this subsection we study the performance of individual formats used for each device. Fig. 7 summarizes the performance on each platform, for each format tested in our analysis separately. Behind each boxplot the height of the bar shows the percentage of matrices in which the specific format exhibited the best performance ('wins'), e.g. for the AMD-EPYC-24 system, MKL-IE was the best format for ∼17% of the dataset. From this figure, there is no evident winner storage format for each device as we observe that all formats are able to achieve a substantial percentage of wins.

We also notice a pattern for the 'research' formats and implementations. Although they can have a formidable number of wins, they usually do not achieve high raw performance. We observe the following: *SparseX* [28] tries to reduce the memory footprint by identifying sub-patterns in the sparse matrix, and therefore mainly targets matrices that do not fit the CPU cache. *SELL-C-σ* [27] tries to take advantage of hardware capabilities like vector extensions in order to improve ILP, so (like all ELL variants) it has an affinity for regular matrices and can compress big sized ones more efficiently. *CSR5* [20] was mainly designed to alleviate the effects of imbalance and irregularity in matrices for GPU architectures. Finally, *Merge-*

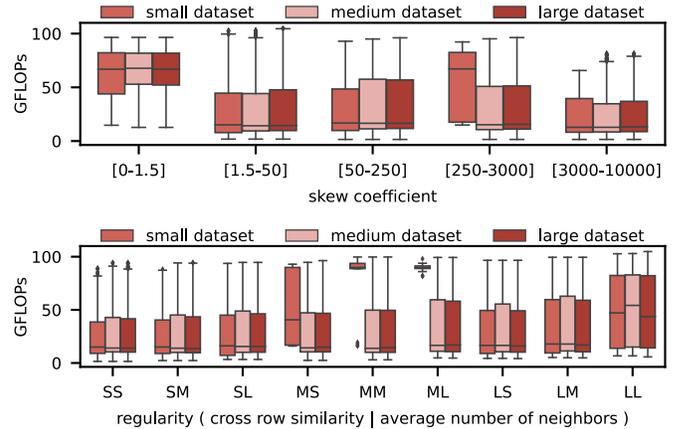

Fig. 8: Comparison of performance variation for three different artificial dataset sizes, tested on the AMD-EPYC-24 CPU. The 'small', 'medium', 'large' datasets consist of 3K, 16K, 27K matrices accordingly.

*CSR* [26] performs well for highly unbalanced matrices, thanks to its 'equal-workload' sharing among threads.

Takeaway 6: *No format/library is a clear winner, and SpMV optimization remains an open field.*

Takeaway 7: *'Research' formats are able to handle cases where the more generic vendor-provided formats cannot perform well, such as large size matrices with high degree of imbalance and irregularity.*

### E. Selection of the artificial dataset size

In this subsection we will discuss how we selected a proper size for the artificial matrix dataset. Our initial evaluation was performed on a small dataset, consisting of approximately 3000 matrices, to resemble the total size of the SuiteSparse sparse matrix collection [30]. However, on this small dataset, we have observed unexpected behavior for specific feature values, as indicatively shown in Fig. 8. We then expanded the 'small' dataset to a 'medium' dataset, while maintaining the feature space limits. This dataset, described in Table I, is the one used in Section V-C for our analysis on all devices. We further expanded the dataset to a 'large' dataset, again by maintaining the feature space limits and sampling more feature values. The 'large' dataset, consisting of 27000 matrices, was tested on the AMD-EPYC-24 testbed, to determine whether more matrices would introduce a fresh perspective on our observations. As shown in Fig. 8, the magnification of the dataset did not alter the overall performance trend, compared to the 'medium' dataset. Given that the process of benchmarking on all devices is time-intensive and power-hungry, we concluded that the 'medium' dataset consisting of 16K matrices has the appropriate size to study the performance of SpMV across all devices.

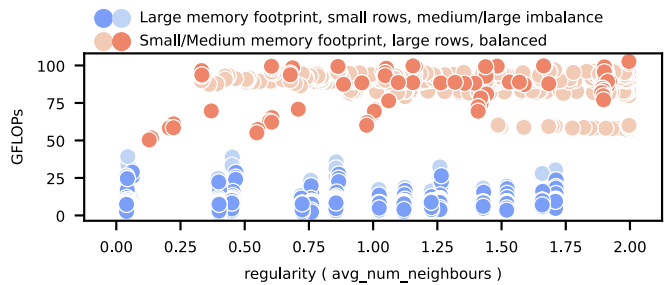

Fig. 9: Evolution of performance on the AMD-EPYC-24 CPU, when the regularity feature grows. Different colors represent different combinations of 'fixed' feature value ranges (memory footprint, row length, imbalance).

It is also necessary to highlight the generator's flexibility; a user can generate matrices on a specific feature space, when research needs to focus on specific matrix properties or a specific architecture.

### F. Opportunities for a more in-depth data analysis

We finally examine how the collected results can be utilized to better understand the effect of a specific bottleneck on each device. In Fig. 9, for the AMD-EPYC-24 device, the average number of neighbors subfeature (from the regularity feature) is allowed to vary, and the other three features are split and kept fixed in three value ranges; small, medium, large. We note that, when a matrix has intuitively "bad" values for CPU execution for the three 'fixed' features, low performance is inevitable, whichever value of regularity may be examined. In this case, only 40% of peak performance can be achieved at most. On the other hand, for intuitively good fixed features (small- or medium-sized matrices, with large rows and low imbalance), increasing the average number of neighbours per row can indeed improve performance by up to $1.6\times$. For the sake of brevity, we do not provide any additional plots. Similar performance studies can be performed for the other features and devices, in order to determine the optimal feature value ranges per device.

## VI. RELATED WORK

Sparse Matrix-Vector Multiplication is a very well studied kernel. Early work on SpMV includes the proposition of storage formats, with COO and CSR/CSC being agnostic to the sparsity structure of the matrix, and diagonal (DIA, JAD) or blocked formats (BCOO, BCSR, VBR) representing specific structures.

In practice, SpMV performance for different formats depends both on the sparsity pattern of the matrix and the architectural features of the target system [33].

Much effort has been put to optimize the serial SpMV [34], [35], focusing on register-level and cache-level blocking. Early works on multicore CPUs [1], [23] identify the bottlenecks discussed in this work, and propose techniques to alleviate them through autotuning [36], and compressed formats [28], [37], [38]. Later works propose formats for manycore processors [27], [39], [40], focusing on SIMD efficiency. The state-of-practice in SpMV on multicore CPUs includes CSR SpMV implementation and the Inspector-Executor CSR SpMV of the Intel MKL library [41], the CSR SpMV and a matrix-structure-optimized SpMV in ARM Performance Libraries [42] and the Inspector-Executor CSR SpMV of the AMD AOCL library [43].

SpMV optimization on GPUs has followed a similar trend, with different formats targeting different bottlenecks, namely HYB [18] targets SIMD optimizations and load balance, CSR5 [20] also targets high SIMD utilization, Merge [26] targets load balance, and BCCOO [19] addresses memory limitations. An exhaustive survey of methods for SpMV on GPUs [21], highlights the absence of a single format to achieve optimal performance for all matrices. The state-of-practice in SpMV for GPUs includes the NVIDIA cuSPARSE library [22], which supports multiple storage formats and the corresponding SpMV functions, including blocked CSR formats.

FPGA designs have focused on CSR customizations [44]–[49], in attempts to address the disadvantage of FPGAs when it comes to memory bandwidth. Recently, commercial FPGAs have been equipped with high bandwidth memory (HBM) channels. Recent works [50]–[52] propose HBM-based SpMV accelerators, however they are designed to support fixed-point or single floating-point precision. The state-of-practice in SpMV for FPGAs includes the Vitis Sparse Library [25], an HBM-oriented CSC SpMV accelerator.

Numerous works focus on predicting the optimal storage format for SpMV on multicore and manycore CPUs and GPUs, based on the input matrix [3]–[11] and autotuning SpMV on the target platforms [12]–[17]. The artificial matrix generator described in our work is inspired by a subset of these works [4]–[6], [8], [13]–[15], [17], which define matrix features and explore or model their relevance to the expected performance, with respect to the storage format.

While all research works related to the SpMV perform performance and/or energy efficiency evaluation, recent studies exclusively focusing on the performance of SpMV are limited [53], [54]. To the best of our knowledge, this is the first comprehensive study of the performance and energy efficiency of the state-of-practice in Sparse Matrix-Vector Multiplication on a variety of modern architectures. We argue that the updates on the performance behavior of SpMV are necessary to support the evolving efforts in optimizing this highly critical kernel.

Research on sparse matrix applications is typically enabled by sparse matrix collections [29], [30], as randomly generated matrices do not represent real sparse matrices in terms of per-formance and numerical stability. Our sparse matrix generator is designed to provide sparsity patterns that imitate real-world sparse matrices, to enable research with respect to the sparsity pattern of the matrix on different architectures. We note that we do not consider numerical aspects of sparse matrices.

## VII. CONCLUSIONS

In this work, we perform an extensive study of the behavior of the Sparse Matrix-Vector Multiplication on modern architectures. Our work is driven by known bottlenecks of SpMV, namely memory bandwidth intensity, low ILP, load imbalance and memory latency overheads, which can appear on all architectures depending on the sparsity pattern of the matrix, the selected storage format and the respective algorithm. We garner five features of the sparsity pattern of the matrix which we perceive as the most influential to performance. Using a sparse matrix generator, based on these features, we create a dataset of 16200 matrices, which we validate against real matrices as to their expected performance. With this extensive dataset at hand, we explore the performance and energy efficiency of SpMV on nine different modern architectures (two AMD EPYC, an ARM NEON, an Intel Xeon and an IBM Power9 CPUs, NVIDIA P100/V100/A100 GPUs, Xilinx Alveo U280 FPGA), using 'state-of-practice' and 'research' storage formats and SpMV implementations.

Our findings shed light on the potential of SpMV execution on contemporary architectures. We conclude that, although GPUs maintain their dominant role in high-performing SpMV, CPUs are very competitive in the execution of small and medium matrices, while FPGAs constitute the most energy-efficient option. Through our detailed feature exploration, we quantify the impact of each bottleneck on the various architectures. The size of the matrix dominates the expected performance, followed by the average number of nonzeros per row, indicating that SpMV remains a memory-bound algorithm, but bottlenecks like low ILP and even lower operational intensity that show up in matrices with small rows are also major concerns. Memory latency overheads due to irregular memory accesses are mostly pronounced on GPUs. Load imbalance, on the other hand, is effectively handled by most storage formats, despite the increased parallelism of modern architectures.

## ACKNOWLEDGMENTS

We thank our anonymous reviewers for their insightful feedback and suggestions on the paper. We also thank the members of the CSLab Research Group for their valuable comments on this work. This work has received funding from the European Union's Horizon 2020 research and innovation programme and by the Greek General Secretariat for Research and Innovation of the Ministry of Development and Investments under the EuroHPC JU grant agreements No. 956416 (project exaFOAM) and No. 955739 (project OPTIMA).